# Diversity of Parahydrogen-Induced Hyperpolarization Effects in Chemistry

Andrey N. Pravdivtsev*[a], Ben. J. Tickner[b], Stefan Glöggler[c,d,e], Jan-Bernd Hövener[a], Gerd Buntkowsky[f], Simon B. Duckett[b], Clifford R. Bowers[g], Vladimir V. Zhivonitko*[h]

| | |
|---|---|
| [a] | Department Section Biomedical Imaging, Molecular Imaging North Competence Center (MOIN CC), Department of Radiology and Neuroradiology, University Medical Center Kiel, Kiel University, Am Botanischen Garten 14, 24118, Kiel, Germany, E-mail: andrey.pravdivtsev@rad.uni-kiel.de |
| [b] | Centre for Hyperpolarization in Magnetic Resonance (CHyM), University of York, Heslington YO10 5NY, UK |
| [c] | Max-Planck-Institute for Multidisciplinary Sciences, Am Faßberg 11, 37077 Göttingen, Germany |
| [d] | Center for Biostructural Imaging of Neurodegeneration (BIN), Von-Siebold-Str. 3a, 37075 Göttingen, Germany |
| [e] | Advanced Imaging Research Center, The University of Texas Southwestern Medical Center, Dallas, Texas 75390, United States |
| [f] | Eduard-Zintl-Institut für Anorganische und Physikalische Chemie, Technische Universität Darmstadt, Peter-Grünberg-Str. 8, D-64287 Darmstadt, Germany |
| [g] | Department of Chemistry and National High Magnetic Field Laboratory, University of Florida, Gainesville, FL 32611, United States |
| [h] | NMR Research Unit, University of Oulu, P.O. Box 3000, Oulu, 90014, Finland, E-mail: vladimir.zhivonitko@oulu.fi |

**Abstract**

Nuclear spin hyperpolarization utilizing parahydrogen has the potential for broad applications in chemistry, biochemistry, and medicine. This review examines recent chemical and biochemical insights gained using parahydrogen-induced polarization (PHIP). We begin with photo-induced PHIP, which allows the investigation of short-lived and photo-activated catalysis. Next, we review the partially negative line effect, in which distinctive lineshape helps to reveal information about rapid exchange with parahydrogen and the role of short-lived catalytic species. The NMR signal enhancement of a single proton in oneH-PHIP is discussed, challenging the underpinning concept of the necessity of pairwise hydrogenation. Furthermore, we examine metal-free PHIP facilitated by novel molecular tweezers and radicaloids, demonstrating alternative routes to conventional hydrogenation using metal-based catalysts. Although symmetric molecules incorporating parahydrogen are NMR silent, we showcase methods that reveal hyperpolarized states through post-hydrogenation reactions. We discuss chemical exchange processes that mediate polarization transfer between parahydrogen and a molecular target, expanding the reach of PHIP without synthesizing specialized precursors. We conclude this review by highlighting the role of PHIP in uncovering the $H_2$ activation mechanisms of hydrogenases. By providing a detailed review of these diverse phenomena, we aim to familiarize the reader with the versatility of PHIP and its potential applications for mechanistic studies and chemical analysis.

# 1. Introduction

Without hesitation, nuclear magnetic resonance (NMR) is one of the leading and most widely used spectroscopic methods in chemistry. This success is due to its broad analytical capabilities and non-invasive nature, enabled by the minuscule magnetic moments associated with the nuclear spin of atomic nuclei. Despite its ubiquitous role in molecular characterization and analysis, NMR suffers from low sensitivity, as only a tiny fraction of all available nuclear spins effectively contribute to the total NMR signal. Consequently, combinations of time-consuming signal averaging, concentrated samples (> mM), and elaborate equipment/hardware are often necessary. Addressing these challenges is the focus of much attention, and the scope of NMR applications is constantly expanding through the introduction of more sensitive NMR instruments with higher magnetic fields,[1] cryoprobes,[2] ultrafast sequences,[3] and nuclear spin hyperpolarization methods.[4]

In this review, we focus on using one of many hyperpolarization methods that increase nuclear spin polarization, using various physical and chemical effects. Such processes create transient non-Boltzmann population distributions across closely spaced nuclear spin energy levels. This effect is highly beneficial as NMR signal enhancements of several orders of magnitude can be obtained. For example, at a clinical field strength of 1.5 T, only 0.000125%, about one in 800'000 $^{13}C$ nuclei, is effectively detected (thermally polarized at ambient conditions). When the system is fully polarized, all $^{13}C$ spins contribute to the signal, enhancing the signal—and thus the sensitivity—by more than five orders of magnitude. Hence, hyperpolarization is a versatile tool that aids the magnetic resonance detection of molecules at low concentrations or short lifetime, with diverse applications in biomedical imaging,[5–7] protein studies,[8,9] composition analysis[10,11] and catalysis.[12–14]

Parahydrogen-induced polarization (PHIP), introduced nearly four decades ago,[15–17] exploits the alignment of the nuclear spins in the singlet state spin isomer of molecular hydrogen, parahydrogen ($pH_2$).[18] $pH_2$ reflects the lowest energy spin isomer of dihydrogen and can be easily prepared in an almost pure state by cooling hydrogen gas to 20 K or below, although 77 K is often used for simplicity and yields a $pH_2$ enrichment of about 50%.[18,19] Para-enriched $H_2$ gas can typically be stored for weeks at ambient temperatures and used on demand to generate nuclear spin hyperpolarization. However, as the para-state is perfectly ordered, it does not yield any NMR signal and is considered 'NMR silent'; further chemical reactivity is required to obtain NMR signal enhancements.

In classical PHIP experiments,[15–17] which are also termed hydrogenative PHIP (hPHIP), the two protons of $pH_2$ are typically added at vicinal positions to double or triple CC bonds in an unsaturated substrate through catalytic hydrogenation. As the spin states in the newly formed molecule evolve, the $^1H$ NMR signal can be significantly enhanced. In this regard, two experimental schemes are defined: parahydrogen and synthesis allow dramatically enhanced nuclear alignment (PASADENA), where the hydrogenation process takes place at a high magnetic field,[16] and adiabatic longitudinal transport after dissociation engenders net alignment (ALTADENA), where hydrogenation occurs at low field[20] (**Figure 1**). In both cases, the resultant $^1H$ hyperpolarization can be used as is or transferred to another nucleus, such as $^{13}C$ and $^{15}N$, e.g., for observation to benefit from longer relaxation times, no background signals, and greater chemical shift dispersion relative to $^1H$. While hPHIP has been most commonly achieved by homogeneous Rh or Ru catalysts,[21–24] other metals[25–30] and heterogeneous catalysts have also exhibited PHIP,[31,32] although typically with smaller polarization than homogeneous systems.

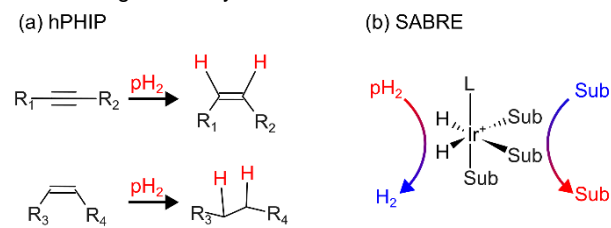

**Figure 1. hPHIP and SABRE.** Schematics of (a) hPHIP, whereby parahydrogen ($pH_2$) is added to an unsaturated bond, and (b) SABRE, where a ligating substrate (Sub) and $pH_2$ undergo reversible exchange at a catalyst. The typical SABRE ligand L is IMes = 1,3-Bis(2,4,6-trimethylphenyl)-1,3-dihydro-2H-imidazol-2-ylidene.

Cis-vicinal hydrogenation is the most typical type of hydrogenation achieved by homogeneous metal catalysts and, therefore, is the most common form observed in hPHIP (**Figure 1a**). This finding reflects the fact that trans-vicinal hydrogenation is unusual in homogeneous catalysis. Recently, trans-vicinal hydrogenation was turned into a dominant pathway to unlock direct hyperpolarization of fumarate[33,34] – an intermediate in the citric acid cycle – and a marker of cell necrosis.[35] In contrast, heterogeneous catalysts (e.g., supported metal nanoparticles) typically yield low hydrogenation selectivity as the reaction mechanism can involve multiple metal sites that deliver their hydrogen atoms to various positions in the product, either directly or reversibly.[13,32] The convenience of using heterogeneous catalysts is now facilitating the production of hyperpolarised gases for use as $pH_2$-enhanced contrast agents in lung MRI.[36,37] In general, hPHIP experiments can also be used to quantify the precise stereo and regioselectivity in such processes through the signal amplification effect.[38–41] Regardless of the hydrogenation mechanism, a crucial requirement to observe hPHIP is the pairwise addition of $H_2$ to a metal catalyst or unsaturated precursor. In other words, both $^1H$ spins of the same $pH_2$ molecule must end up in the same product molecule to retain their spin correlation. This requirement has provided significant insight into metal-based oxidative addition reactions.[42] An alternative to hPHIP is non-hydrogenative PHIP (nhPHIP, **Figure 2**), with its most popular form called signal amplification by reversible exchange (SABRE, **Figure 1b**).[43] In SABRE, $pH_2$ and a substrate ligate transiently with a metal center. Depending on the experimental conditions, J-couplings or cross-relaxation can then drive polarization transfer from the $pH_2$-derived spins into those of the transiently bound substrate in the resulting complex. Subsequent dissociation of this ligand results in a chemically unaltered but hyperpolarized free substrate in solution,[44] allowing for multiple and continuous polarization of the same molecule.[45–47]

For molecules to become hyperpolarized by PHIP, several specific requirements must be met. For hPHIP these include the presence of an unsaturated functionality (typically alkene or alkyne groups) that can accept $pH_2$. The strategy to hyperpolarize molecules that do not possess an appropriate unsaturated bond[48–54] or ligating functionality[55,56] is to add such a function on a sidearm. The resulting side arm is then released after hydrogenation and hyperpolarization transfer by hydrolysis (e.g., PHIP-SAH).[48–54] For SABRE, the substrate and $pH_2$ must ligate to a metal center, form spin-spin interactions, and dissociate. Therefore, such substrates often contain an electron-donating nitrogen site.[43,57] However, the ongoing design of new catalysts has significantly expanded the types of molecules amenable to SABRE to include biologically significant O-donor ketoacids.[58]

Other studies have seen sulfur, phosphorus, and silicon donor sites employed.[59–61]

| | | Type of chemical reaction used to generate PHIP | |
|---|---|---|---|
| | | Catalytic hydrogenation of an unsaturated substrate with pH$_2$ | Transient interaction of a substrate with pH$_2$ on the catalyst |
| Substrate-pH$_2$-relationship | Pairwise pH$_2$ addition | hPHIP (including PHIP-SAH), Photo-PHIP (hydrogenation of substrate) | Reversible exchange of two H atoms in the catalyst (Photo-PHIP, MF-PHIP, SABRE, Photo-SABRE), Pairwise replacement of two H atoms in the substrate with pH$_2$, PNL |
| | Non-pairwise pH$_2$ addition | oneH-PHIP, MF-PHIP, NEPTUN | SWAMP* |
| | Relayed polarization transfer | PHIP-X (also EX, RELAY) | SABRE-Relay |

**Figure 2.** Classification of chemical effects that lead to PHIP. The diagram is inspired by the diagram of Emondts et al. [89]. The primary scope of this review is to describe the pH$_2$-derived hyperpolarization effects except those of routine hPHIP and SABRE, which are well-reviewed. *Preliminary assignment based on current data.

Collectively, hPHIP,[36,62–67] PHIP-SAH,[68–70] and SABRE[71,72] have led to *in vivo* metabolic imaging applications. The translation from optimization of PHIP to *in vitro* and *in vivo* studies is not straightforward but has accelerated in recent years, demonstrating PHIP is a viable route to produce preclinical and clinical hyperpolarized agents in liquid and gas phases competitive with other approaches such as dissolution dynamic nuclear polarization (dDNP)[68] or spin-exchange optical pumping (SEOP)[73,74]. Advances in PHIP for *in vivo* imaging have been well-reviewed.[4,75–77] In addition to these relatively well-known protocols, there are other effects that, although less popular, have unique properties and can provide valuable insight into chemical and biochemical reactivity.

For example, one less common type of hPHIP is geminal hydrogenation, where two protons of pH$_2$ bind to the same carbon.[78,79] There are other variations of the PHIP effect, such as when the catalyst itself interacts with pH$_2$ and becomes transiently polarized.[80–84] Other cases include pairwise replacement of two substrate protons with a proton pair from pH$_2$ with no net hydrogenation,[14,38,85,86] the addition of only one proton from a pH$_2$ molecule to the substrate (oneH-PHIP)[87] (after the formation of an intermediate by pairwise addition) or even hydrogenation accompanied by oligomerization.[88]

This review will focus on a selection of these more unusual PHIP effects beyond PASADENA, ALTADENA, hPHIP, and SABRE. We will discuss these phenomena and their mechanisms (**Figure 2**), seeking to promote new analytical applications and provide valuable insight into the underlying chemical interactions. In this regard, we will discuss photo-induced PHIP and SABRE, partially negative line (PNL) effects and their implications for the analysis of short-lived intermediates, oneH-PHIP effects in hydroformylation, the various mechanisms of hyperpolarization of water using pH$_2$, PHIP effects in metal-free hydrogenation (MF-PHIP) reactions, secondary transformation which reveal hidden PHIP, chemically relayed polarization transfer, and PHIP in enzymatically catalyzed hydrogenation reactions.

## 2. Reviewed variants of PHIP

### 2.1. Photo-PHIP and photo-SABRE

Breaking the bond in pH$_2$ is a critical step in most PHIP-type experiments as it unlocks the spin order of pH$_2$, ideally without losing spin alignment. This step is most commonly associated with its oxidative addition to a metal center. Such metal complexes must, therefore, have appropriate diamagnetic electron configurations that support the formation of two new metal-hydride bonds. Accordingly, many stable complexes must undergo ligand loss before such a process occurs. This has given rise to studies where light irradiation stimulates ligand loss from a stable metal complex to generate a reactive intermediate that can react rapidly with pH$_2$ (**Figure 3a**).[90–96] Consequently, hyperpolarized metal complexes are formed in a process that is somewhat analogous to hPHIP, which involves pH$_2$ addition to a stable 16-electron transition metal precursor. However, combining photochemistry with PHIP, termed photo-PHIP, can create magnetic states that differ from those created under PASADENA, where time averaging eliminates transverse components, but preserves the longitudinal two spin order term associated with the detection of antiphase signals.[16,97,98] This photochemical approach has been used to study rapid kinetics[91,94] and detect reaction intermediates or products.[90,94,99] The examples of photo-PHIP typically involve Ir and Ru carbonyl precursors containing phosphine, diphosphine, or diarsine ligands.[95,96,99,100]

For reference, pH$_2$ exists as a nuclear spin singlet state that is defined by the density matrix described in equation 1

$$\hat{\rho}_{pH_2} = \frac{\hat{1}}{4} - (\hat{I}_x\hat{S}_x + \hat{I}_y\hat{S}_y + \hat{I}_z\hat{S}_z), \quad (1)$$

where $\hat{I}_i$ and $\hat{S}_i$ ($i = x, y, z$) are spin operators which refer to identical atoms in pH$_2$. This state consists of the longitudinal two-spin order $\hat{I}_z\hat{S}_z$, and the in-phase zero-quantum coherence (ZQC) $\hat{I}_x\hat{S}_x + \hat{I}_y\hat{S}_y$. In hPHIP, numerous catalytic hydrogenation steps, or reversible H$_2$ exchange in SABRE, commonly create weakly-coupled metal-dihydrides. In this case, the in-phase ZQC oscillates due to the periodic transformation into the out-of-phase ZQC state, $\hat{I}_x\hat{S}_x - \hat{I}_x\hat{S}_y$. As a result, it averages to zero over the time of thermally initiated reactions,[101] unless strong proton decoupling is applied[65,102] or experiments are performed at low magnetic fields when metal-dihydrides are strongly coupled.[103] For two weakly coupled spins, the oscillation of ZQC occurs at a frequency equal to the difference between the Larmor frequencies of the two nuclei, $v_{ZQC} = |v_I - v_S|$, where $v_I$ and $v_S$ are the Larmor frequencies of the respective spins $I$ and $S$. Typically, in such systems, there are strong interactions between hydride protons and $^{31}$P of spectator phosphine ligands, which additionally contribute to $v_{ZQC}$.

In contrast, the photo-PHIP approach involves short laser-induced photodissociation (on the order of nanoseconds) to create a vacant ligand site such that rapid reaction with pH$_2$ is possible (**Figure 3a**). As a consequence of the relatively facile pH$_2$ addition (faster than $1/v_{ZQC}$ such that ZQC does not have enough time to oscillate) all terms in eq. 1 are preserved through the hydrogenation reaction even though the singlet state is not stationary in the dihydride product complex.[96] Due to rapid $pH_2$ addition and the nonstationary nature of the singlet state, the ZQC component oscillates (**Figure 3b**).

For symmetry reasons, singlet states cannot be perturbed by the hard pulses typically used for NMR excitation, and they consequently do not lead to any observable signals. Therefore, selective excitation of one of the chemically inequivalent ligands is required to observe this magnetization immediately after the laser pulse irradiates the sample at the high-magnetic field of an NMR spectrometer.[104] Alternatively, adiabatic RF pulses can convert the singlet spin order into observable magnetization of one or two protons.[105,106]

As the ZQC term evolves under the chemical shift operator, the new state of the system is NMR observable. It can be probed directly with broad-band pulses (typically 45° or 90°, **Figure 3b**), and its amplitude and phase oscillate with time.[92,102,106] This approach was used to study the evolution of dihydride singlet order for a wide range of Ru and Ir complexes with a variable delay between laser irradiation and NMR detection.[92] This type of laser-pump NMR-probe experiment, over a microsecond to millisecond delay timescale, allowed observation of reactivity in an approach analogous to other laser-based UV and IR time-resolved spectroscopies, which was only possible by marrying laser-induced ligand dissociation with a $pH_2$ addition step (**Figure 3b**).[90–96] Multiple laser pulses or continuous wave laser irradiation over times much longer than $1/v_{ZQC}$ have also been employed.[94,96] However, in these cases of slow hyperpolarization preparation, spin order is averaged across the light irradiation time, which leads to an effect analogous to the time-averaged PASADENA effect as the ZQC terms are lost (**Figure 3c**).

The analysis process required to extract information about the kinetics of $pH_2$ addition in these experiments is complex.[91,93,106] For example, in cases where the $pH_2$ addition rate is on the same order of magnitude as the frequency of ZQC evolution, $v_{ZQC}$, only partial averaging occurs. This can lead to apparent phase shifts in the signal oscillation that contain quantitative information about the $pH_2$ addition rate.[94] However, if product formation is much faster than the rate of ZQC evolution, then only an upper bound for the $H_2$ addition rate can be determined. On the other hand, if product formation is much slower than the characteristic frequencies of the spin system evolution, the ZQC oscillations are no longer observed due to complete averaging. The latter case is similar to traditional PHIP in which thermally controlled reactions (rather than laser-induced) build up the number of $H_2$ addition products over a longer time window (ca. seconds).[91] To this end, the photo-PHIP method has been applied to measure the $H_2$ addition rate of $[Ir(I)(PPh_3)_2(CO)]$, which is formed from laser-induced $H_2$ dissociation from $[Ir(I)(H)_2(PPh_3)_2(CO)]$ (**Figure 3b**).[94] Notably, the obtained rate constants were comparable to those measured using flash photolysis coupled with optical spectroscopy.[94] However, unlike flash photolysis, NMR has additional chemical resolution that enables a more detailed analysis of chemical reactions, which is especially important for mixtures of photoactive substrates.

Approaches of this type have also been used to study ruthenium arsine complexes that act as alkyne hydrogenation catalysts as catalytic activity is observed after photolysis, and many intermediates and species involved in the catalytic cycle are detected and characterized with the help of photo-PHIP-enhanced signals.[95] Due to the lower reaction rate of these catalysts, these types of examples can be called slow photo-PHIP. The delay between laser irradiation and NMR signal acquisition is much longer (on the order of seconds), and ZQC completely decays, revealing enhanced antiphase NMR signals analogous to those in traditional PASADENA experiments without amplitude oscillations.

Initiating hydrogenation catalysis with a laser pulse allows access to hyperpolarized species inaccessible in thermal reactions (without laser-induced dissociation), as thermal conditions are insufficient to enable ligand dissociation, which must occur before $pH_2$ addition for many metal complexes. One recent example also involved the use of an irradiated iridium photosensitizer to produce excited $[Ru(H)_2(PPh_3)_3(CO)]$, which in turn stimulated $H_2$ dissociation to form $[Ru(PPh_3)_3(CO)]$. Subsequent hydrogenation of phenylacetylene with $pH_2$ yielded PHIP-enhanced styrene (single hydrogenation) and ethylbenzene (double hydrogenation) with $^1H$ NMR signals for these organic photoactivated hydrogenation products enhanced by up to 1630 times at 9.4 T (**Figure 3c**).[107]

While there are several examples of photoactivated PHIP catalysts, we are unaware of examples of photoactivated SABRE catalysts. So far, SABRE typically relies on iridium catalysts that reversibly exchange $H_2$ under thermal conditions. Rational catalyst design is, however, often used to improve the efficiency of SABRE by tuning ligand exchange rates or controlling the binding of particular target substrate molecules.[14]

One example of a type of photo-SABRE involved azobenzene, with light irradiation of the target ligand rather than the metal catalyst. This approach used light irradiation to switch azobenzene between *cis* and *trans* conformations, controlling SABRE activity as only *cis*-azobenzene has appropriate geometry for ligation to the iridium SABRE catalyst (**Figure 3d**).[108] Upon hyperpolarization of its $^{15}N$ nuclei using a traditional SABRE approach, light-induced isomerization locked the polarization within *trans*-azobenzene as it does not interact with the metal center, allowing for prolonged polarization lifetimes. Yet another benefit was that the *trans* isomer also forms a long-lived spin state with a lifetime of *ca.* 25 minutes in which polarization can be stored. While this example utilized a photo-switchable target, future examples of light-controlled SABRE *via* photosensitive catalysts with light-initiated $H_2$ or substrate exchange may arise from new SABRE catalysts.

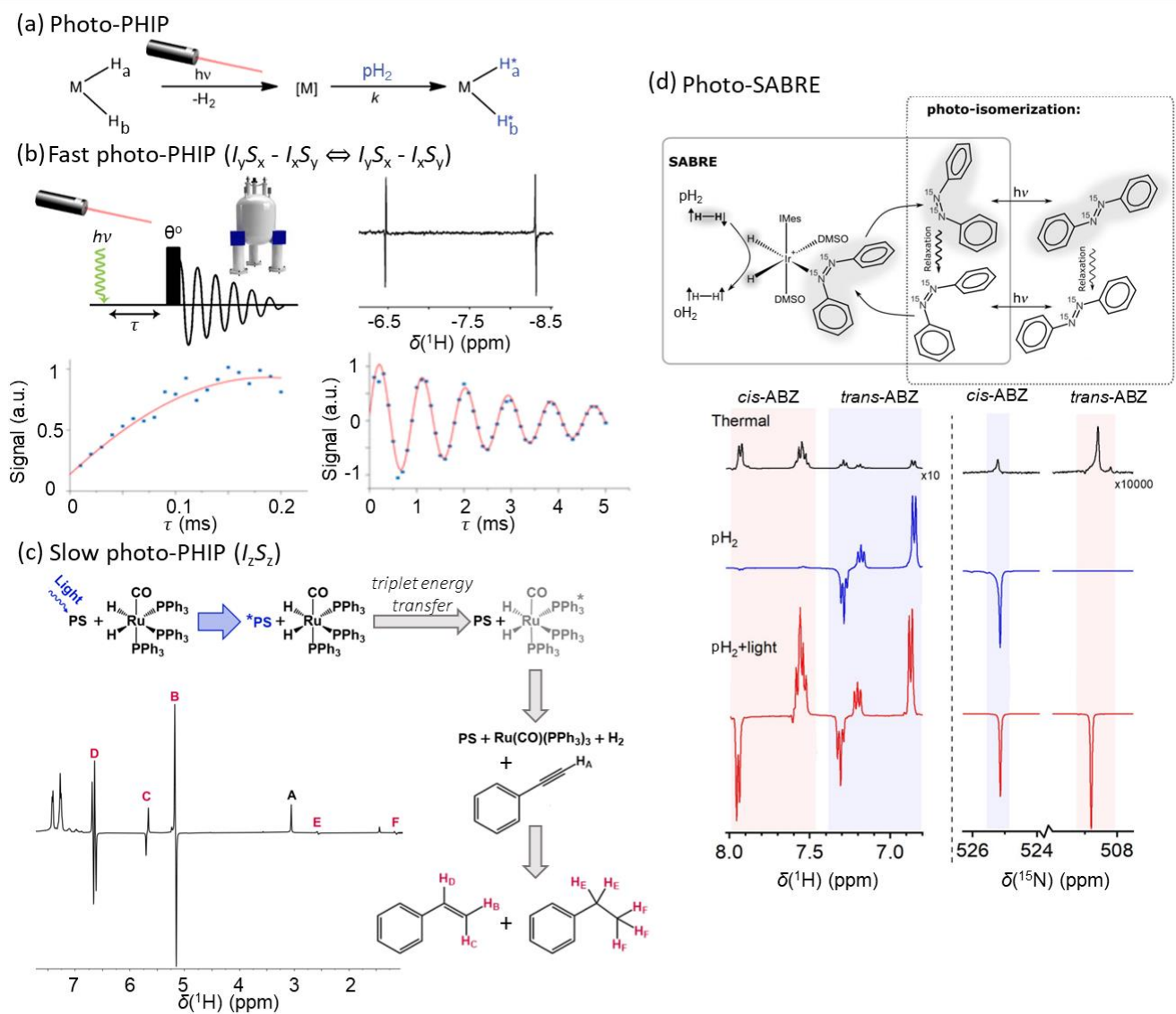

**Figure 3. Examples of photo-PHIP.** (a) The outline of a photo-PHIP experiment in which light-induced ligand dissociation creates a metal complex [M] that can add pH$_2$ to create an inequivalent metal dihydride species with PHIP-enhanced hydride NMR signals. (b) Depiction of laser-pump NMR-probe spectroscopy in which laser irradiation stimulated H$_2$ – pH$_2$ exchange followed by NMR detection (where θ is typically 45° or 90°), with a variable delay between the two (top left). The NMR spectrum at the top right is a single-laser shot, single 90° RF pulse, pH$_2$-enhanced $^1$H{$^{31}$P} NMR spectrum of the hydride region of [Ru(PPh$_3$)$_3$(CO)(H)$_2$] with τ = 0.05 ms (broadband $^{31}$P decoupling). When the delay is on the millisecond timescale, the zero quantum coherences of pH$_2$ in the chemically inequivalent dihydride complex can evolve due to the chemical shift difference from unobservable $\hat{I}_x\hat{S}_x + \hat{I}_y\hat{S}_y$ to observable $\hat{I}_y\hat{S}_x - \hat{I}_x\hat{S}_y$. The lower kinetic traces show the hydride signal integral at δ ~ – 6.5 ppm from $^1$H{$^{31}$P} NMR spectra of [Ru(PPh$_3$)$_3$(CO)(H)$_2$] (1 × 10$^{-3}$ M, 3 bar pH$_2$) as a function of τ. Experimental points are shown in blue with fitted red lines. This spin evolution can give information about kinetic hydrogen addition rates. (c) Slow photo-PHIP leading to hyperpolarized styrene and ethylbenzene after [Ru(H)$_2$(PPh$_3$)$_3$(CO)] and an iridium photosensitizer (PS) are irradiated at 420 nm in DCM-d$_2$ at 298 K.[107] In these cases, delays between laser irradiation and NMR pulse acquisition are on the order of seconds. Consequently, ZQC coherences are not observed, which is similar to the traditional PASADENA effect. (d) Depiction of photo-SABRE in which a SABRE-catalyst hyperpolarizes *cis*-azobenzene (ABZ) followed by light irradiation to switch between *cis* and *trans* isomers of ABZ (upper). Hyperpolarized molecules are shown with a grey background.[108] Example $^1$H (lower left) and $^{15}$N (lower right) NMR spectra were recorded after 10 minutes of light irradiation of a sample containing 56 mM of $^{15}$N$_2$-ABZ in CD$_3$OD with 1 mM of [IrCl(COD(IMes)] and 200 mM DMSO-d$_6$ at 9.4 T. SABRE polarization transfer to $^1$H was performed at 200 nT and to $^{15}$N at =400 nT. Signals for *cis*-ABZ are shown with *red* background, with those of *trans*-ABZ with a blue background. *(a) Adapted with permission from Procacci et al.*[94] *Copyright 2016 Royal Society of Chemistry. This publication is licensed under CC BY-NC 3.0. (b) Adapted from Torres et al.*[93] *Copyright 2014 with permission American Chemical Society. This publication is licensed under CC-BY. (c) Adapted with permission from Brown et al.*[107] *Copyright 2022 The Authors. Published by American Chemical Society. This publication is licensed under CC-BY 4.0. (d) Adapted from Kiryutin et al.*[108] *Copyright 2023 with permission Wiley-VCH GmbH. License number 5850221124446.*

## 2.2. Reversible H$_2$ exchange and the partial negative line (PNL) effect

The effects of intra- or intermolecular chemical exchange can strongly influence the NMR line shapes of molecular complexes interacting with pH$_2$, hydrogenation products, and dissolved dihydrogen (H$_2$). This exchange can act as an additional relaxation mechanism promoting singlet-triplet state transitions and singlet state depletion.[98,109] Consequently, NMR line shape analysis can be employed as an analytical tool to monitor transient states exchanging with hydrogen and/or catalyst, *via* changes in the line shape of the H$_2$ signal.[110] This effect was independently observed as an antiphase line for free-dissolved H$_2$ by Zhivonitko et al.[111] during the investigation of pH$_2$ activation with *ansa*-aminoboranes and by Kiryutin and Sauer et al.[110] during the investigation of the PHIP enhancement of small oligopeptides (**Figure 4**). This phenomenon is now referred to as the partially negative line shape (PNL) of the dihydrogen NMR signal.

Kiryutin et al. showed that this effect is independent of the presence of a substrate and occurs *via* transient interaction of the pH$_2$ molecules with the catalyst under fast chemical exchange. Using the double quantum coherence filter variant of the only parahydrogen spectroscopy (OPSY) NMR pulse sequence,[112–115] the thermal orthohydrogen (oH$_2$) background signal was suppressed, and it was proven that the PNL effect results from a two-spin order state. Typically, the two-spin order term of H$_2$ is NMR invisible. However, the exchange between free H$_2$ and two chemically nonequivalent hydride ligands in a transient catalyst-dihydrogen complex leads to a slight frequency difference for two $^1$H resonance components of free H$_2$. Consequently, when pH$_2$ is used, the singlet spin order evolves on the complex, leading to PASADENA-type exchange broadened antiphase signals of dihydride ligands. In contrast, the H$_2$ signal is revealed due to the exchange dynamics as a superposition of two enhanced resonance lines with opposite phases and slightly shifted frequencies.

Based on this explanation of the PNL effect, the partially negative line experiment (PANEL)[110] was developed (**Figure 5a**), which employs continuous-wave low-power radio frequency irradiation to detect transient species indirectly by observing the change in the PNL signal of free H$_2$. When the frequency of the continuous radiofrequency pulse is in resonance with one of the hydrogens bound to the short-lived catalyst or free hydrogen, this nucleus is saturated, and the PNL is strongly affected. This experiment is similar to chemical exchange saturation transfer (CEST) NMR experiments, which have also been used to study short-lived exchanging intermediates.[116] Using PANEL, a sensitivity gain of at least three orders of magnitude, compared to routine NMR experiments, is achievable (**Figure 5b**). As with CEST, the sensitivity and spectral resolution of PANEL depend on the amplitude of the RF saturation field.[117] Since hydride protons typically resonate at chemical shifts below -10 ppm at high magnetic fields, RF pulses of 1 kHz can be applied without distorting the PNL line of H$_2$, which appears at around 4.5 ppm. Hence, PANEL allows observation of intermediates with lifetimes of 1 ms or even shorter. The achievable spectral resolution of the PANEL depends on the RF power. Resolutions of 1 ppm are achievable by employing RF amplitudes on the order of ca. 100 Hz.

The presence of two peaks of the same intensity at -16.5 ppm and -13.5 ppm in the PANEL spectrum (**Figure 5b**) points to the sizable chemical shift difference between two inequivalent hydrides and the corresponding complex asymmetry. In such a weak coupling case ($|\omega_I - \omega_S|/2\pi \gg |J_{IS}|$) the H$_2$ resonance splitting that leads to PNL is given by equation 2[110]

$$\Delta\nu \approx \frac{K_{eq}k_d^2}{2}J_{IS}\left(\frac{k_d^2-(\omega_I-\omega_H)^2}{(k_d^2+(\omega_I-\omega_H)^2)^2} + \frac{k_d^2-(\omega_S-\omega_H)^2}{(k_d^2+(\omega_S-\omega_H)^2)^2}\right) \quad (2)$$

where $\omega_I$, $\omega_S$, $\omega_H$ are frequencies of the spins in the complex (I and S) and of free dihydrogen (H), $J_{IS}$ is the constant of spin-spin interaction in the complex, $K_{eq} = k_f$ [ML$_n$] / $k_d$ is the equilibrium constant of the binding of dihydrogen to the complex, $k_d$ is the dissociation rate constant, and $k_f$ is the formation rate constant.

The same theoretical model of PNL[110] reproduced the experimental NMR line-shapes, the nutation angle dependence, and the dependence on the frequency of the resonance position of the PNL. It also permitted the determination of chemical shift values for exchanging protons in the transient complex and the sign of the scalar coupling constant between those protons. Typically, hydride resonances of transient complexes are hardly observable directly in $^1$H NMR, whereas PNL allows for indirect but sensitive detection of their presence.[110] In parallel to that, Johnson et al.[118] observed PNL and then managed to detect a hyperpolarized dihydrogen complex by cooling the catalyst sample with pH$_2$ down to 238 K, confirming the presence and rapid exchange of the dihydrogen. The observation of reaction intermediates is highly valuable as such species are difficult to discern in standard $^1$H NMR experiments due to their low signal intensity and broad line widths. These techniques were later applied to study several exchange pathways with pH$_2$.[119]

Bernatowicz et al.[120] proposed an alternative explanation for the creation of PNL, which suggests it is caused by residual dipolar couplings (RDCs) stemming from the partial ordering of the hydrogen molecules in the external magnetic field. However, experimental studies into this mechanism are warranted.

Czarnota et al.[121] and Alam et al.[122] investigated the conditions for a PNL effect employing iridium complexes or metal-organic frameworks (MOFs). PNL effects were found in the catalytic hydrogenation of eptifibatide, a disintegrin derivative based on a protein from the rattlesnake venom,[123] or trivinyl orthoacetate,[79] as well as in SABRE studies employing Zintl cluster supported rhodium centers[124] or nickel diazadiphosphacyclooctane complexes.[119] Finally, PNL effects are also a possible loss channel for hyperpolarization in hPHIP or SABRE applications.[98,125–127] PNL not only allows for ultra-sensitive detection of intermediates but can also serve in the future as a monitor for spin dynamics and thus chemical kinetics in intermediate complexes.

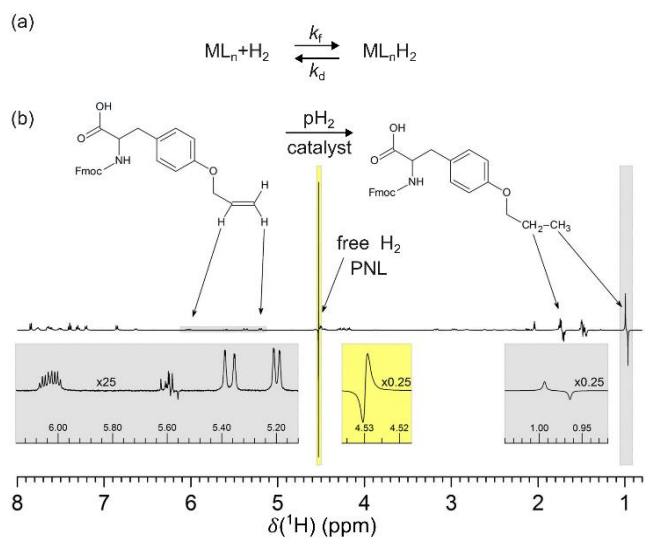

**Figure 4. Partial Negative Line effect induced by a catalyst - pH$_2$ interaction.** (a) Chemical exchange of pH$_2$ with the metal center ML$_n$ of the catalyst results in the PNL effect on free H$_2$. (b) $^1$H PHIP NMR spectrum recorded during the hydrogenation of Fmoc-O-allyl-tyrosine (left) to Fmoc-O-propyl-tyrosine (right) with pH$_2$, showing a strong PNL signal for free H$_2$ at 4.53 ppm. *Adapted from Kiryutin et al.* [110] *Copyright 2017 with permission from American Chemical Society.*

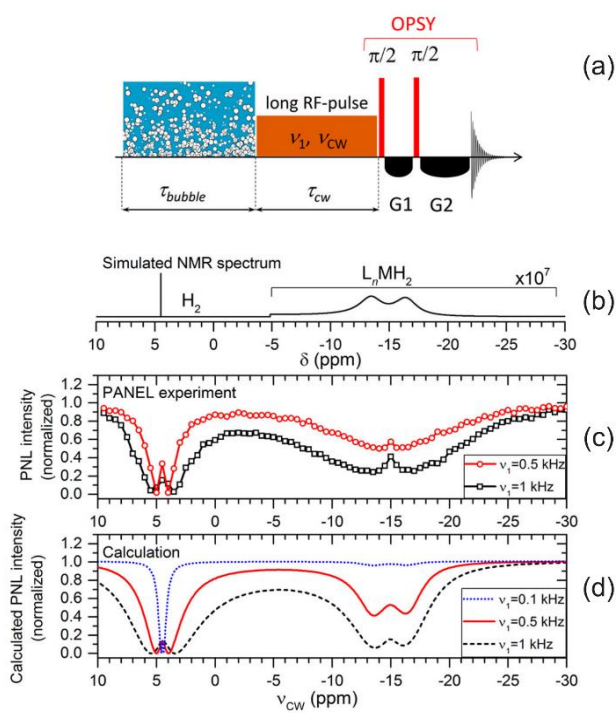

**Figure 5. PANEL – A combination of the CEST method and the PNL effect boosts the sensitivity of intermediate complex evaluation.** (a) Scheme of the PANEL (partially negative line) experiment for indirectly detecting the hidden hydrogen catalyst complex ($L_nMH_2$). (b) Calculated NMR spectrum of the hidden complex, magnified by a factor of $10^7$. (c) Experimental and (d) simulated PANEL spectra showing the signal of the hidden complex (lines at -16.5 ppm and -13.5 ppm). *Adapted from Kiryutin et al.* [110] *Copyright 2017 with permission from American Chemical Society.*

**2.3. OneH-PHIP**

The typical pairwise addition of $H_2$ for hPHIP can be lifted if the final reaction product is derived from an intermediate that is itself formed by pairwise hydrogenation when the lifetime of the intermediate is sufficient to allow the singlet spin order of the $pH_2$-derived proton pair to evolve into Zeeman order (before separation). In such cases, the hyperpolarization observed in the final product can be associated with no $pH_2$-derived protons as in SABRE or with only one of two $pH_2$-derived protons. The latter is referred to as the oneH-PHIP effect.

OneH-PHIP was first observed by Permin and Eisenberg during their stoichiometric studies of hydroformylation by platinum-tin and iridium carbonyl species.[87] Here, *trans*-PtCl(COEt)(PPh$_3$)$_2$ proved to react with SnCl$_2$ and $pH_2$ to form propanal, where the slowly relaxing aldehydic proton (**Figure 6a**) exhibited NMR signal enhancement. This is reflective of the creation of single-spin net polarization associated with an $\hat{I}_z$ type term (indicated by **H** in the following products), which differs significantly from the more usual $pH_2$-derived longitudinal two-spin order $\hat{I}_z\hat{S}_z$ term (the last term in eq. 1), which is destroyed when the coupling between the spins is lost. As this section will illustrate, such effects are relatively common, although the signal enhancements are relatively low. For example, a signal enhancement of 5-fold at 9.4 T using 50% $pH_2$ was achieved for the aldehydic proton of propanal.[87] Hence, oneH-PHIP observation has been limited to slowly relaxing species or high-turnover catalysis.

Permin and Eisenberg investigated the catalytic production of propanal (CH$_3$CH$_2$C**H**O) using PtCl$_2$(CO)(PPh$_3$)-SnCl$_2$ [87] and [Ir(COEt)(CO)$_2$(dppe)] (dppe is 1,2-bis(diphenylphosphino)ethane) wherein only the aldehydic proton was hyperpolarized (**Figure 6a**). In the case of Ir systems, hyperpolarized hydride ligand signals were also detected for the intermediate [Ir(**H**)$_2$(COEt)(CO)(dppe)], where the hydride ligand *cis* to the phosphines (**H'**) was then found to become the hyperpolarized aldehydic proton in the final product. Hydride signals for this intermediate appear at very close resonances (-8.696 and -8.905 ppm in benzene), forming an AB-type spin system. In Pople notation, the AB-type spin system corresponds to two spins exhibiting a chemical shift difference comparable to their spin-spin coupling. The observed hyperpolarization of only a single proton in the final aldehyde product was called oneH-PHIP. Subsequent solvent variation of a benzene-acetone mixture led to an inversion in the oneH-PHIP phase of the aldehydic proton as a consequence of the relative change in chemical shifts of the AB spin system of the hydride ligands inverting. Hence, the oneH-PHIP effect was linked to strong coupling. In this case, a rigorous theoretical description, enunciating the role of chemical shift difference and mutual spin-spin interaction in the AB spin system that created the $\hat{I}_z - \hat{S}_z$ type magnetization was reported, which set Bargon's earlier description into a firm chemical context.[128] However, it does not exclude the possibility of a relaxation-driven polarization transfer mechanism. For instance, in section 2.5, we discuss such mechanisms in the hydrogenation of alkynes and imines with $pH_2$ using *ansa*-aminoborane catalysts.[129–131] To verify the exact mechanism, one can study the magnetic field dependence of the oneH-PHIP effect: cross-correlated relaxation is expected to be stronger at higher fields, while coherent mechanisms dominate at lower fields.[128,132,133] Another rationale for the oneH-PHIP effect appears in Ref.[134], although this time in the context of the hyperpolarization of water and alcohols by either homogeneous or heterogeneous catalysis, as described in Section 2.4.

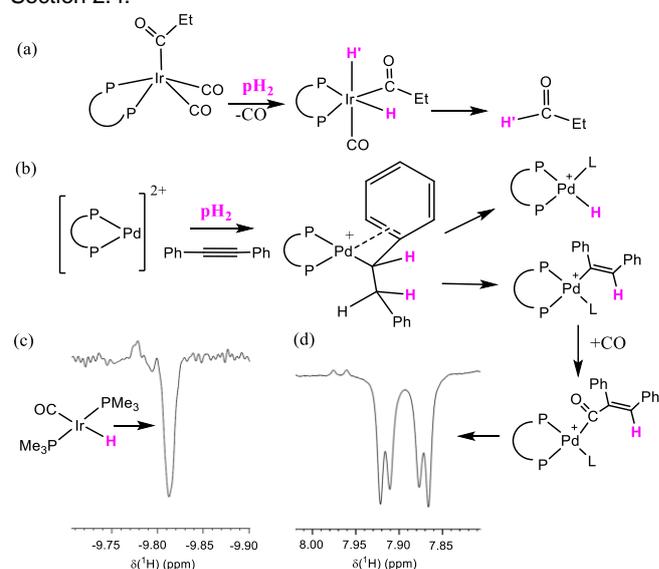

**Figure 6. The oneH-PHIP effect** has allowed the detection of several single-spin hyperpolarized products. This effect results from the creation of a hyperpolarized AB-type spin system for the $pH_2$-derived protons (indicated with H in (a) and (b)), with subsequent further reaction producing single spin hyperpolarized products. Examples shown here include (a) aldehydes[87] (b) and (c) metal hydride complexes[135,136] and (d) vinyl-containing species.[137] The NMR traces illustrate the typical appearance of a single spin-polarized resonance under $^{31}P$ decoupling in the case of (c).

An additional example of a hydride resonance exhibiting oneH-PHIP was observed during catalytic studies of alkyne hydrogenation by [Pd(bcope)(OTf)$_2$] (bcope = (*c*-C$_8$H$_{14}$-1,5)PCH$_2$CH$_2$P(*c*-C$_8$H$_{14}$-1,5); OTf = CF$_3$SO$_2$O$^-$) where species like [Pd(bcope)(pyridine)(**H**)](OTf) were detected.[136] This study extended into the detection of critical reaction intermediates like [Pd(bcope)(CPh=C(**H**)Ph)(pyridine)](OTf), where the single vinyl proton exhibited strong hyperpolarization, alongside cis-Ph**H**=C**H**Ph (**Figure 6b**). During these studies, it was the strongly coupled spin system of the reversibly formed intermediate [Pd(bcope)(CHPhC(H)$_2$Ph)](OTf) that led to this behavior.[136] The related complex, the alkene insertion product, Pd(Ph$_2$PCH$_2$CH$_2$PCy$_2$)(-C(Ph)H–CHPh–CPh=(C**H**)Ph)]OTf has also been observed thanks to the $^1$H NMR signal enhancement of oneH-PHIP[138] and vinyl ethers have been produced during platinum-catalyzed reactions that exhibit this effect.[139]

Furthermore, the addition of CO to drive palladium-catalyzed carbonylation has extended the hyperpolarized observations to include the ketone MeOCO(CPh)=C**H**Ph proton resonance, alongside further signals in the acyl bearing reaction intermediate [Pd(bcope)(CO-CPh=C(**H**)Ph)(CO)](OTf), the novel hydride complex [Pd(bcope)(CO)(**H**)](OTf), the alkene complex [Pd(bcope)(C**H**Ph=CPh(COOMe)] and free **H**D (alongside **H**$_2$) (**Figure 6d**).[137] Here again, detecting a hyperpolarized response for the released H$_2$ reflects the oneH-PHIP that results from strong coupling effects in species that led to it.

Later, Guan et al. reported on studies of related Ir(η$^3$-C$_3$H$_5$)(CO)(PMe$_3$)$_2$ type species and described how the hydride ligand signal for **H**Irl(CO)(PMe$_3$)$_2$ exhibited oneH-PHIP (**Figure 6c**).[135]

It should be apparent from these discussions that the sharing of hyperpolarization between species can occur *via* numerous processes with dramatically different efficiencies. The polarization of a single proton was also reported in other reactions with H$_2$ involving metal-free catalysts or heterogeneous catalysts and led, among others, to a polarization of water, as discussed in the following sections.

## 2.4. SWAMP and NEPTUN

Hyperpolarized water can be used for angiography and perfusion biomedical imaging[140–142] and as a polarization source for heteronuclear signal enhancement in biomolecular NMR spectroscopy.[9,143–146] While high polarization levels (>60%) have been shown to result from dDNP,[147] pH$_2$-based methods offer an alternate route that is both rapid and less expensive, thus making the approach more widely accessible.

Water had been an elusive target for PHIP until 2017, when it was hyperpolarized in D$_2$O mixtures of L-histidine and a water-soluble iridium complex, [Ir(Cl)(IDEG)(COD)] (IDEG=1,3-bis(3,4,5-tris(diethyleneglycol)benzyl)imidazole-2-ylidene).[148] In this system, the hyperpolarized HDO and HD proton signals appear in the emission and absorption phases, respectively (**Figure 7a,b**). Consistent with the oneH-PHIP theory,[89,134] the enhanced HD and HDO signals have opposite phases and very similar field dependencies of signal amplitudes, which reach a maximum near 45 mT where the *J*-coupling and chemical shift difference between the dihydride protons (**Figure 7c**) are matched. The corresponding oneH-PHIP mechanism, mediated by (*i*) H/D exchange with coordinated D$_2$O, (*ii*) dissociation of HDO, and (*iii*) H-D recombination, was named nuclear exchange polarization by transposing unattached nuclei (NEPTUN).[89] The role of L-histidine in this work remains unclear.

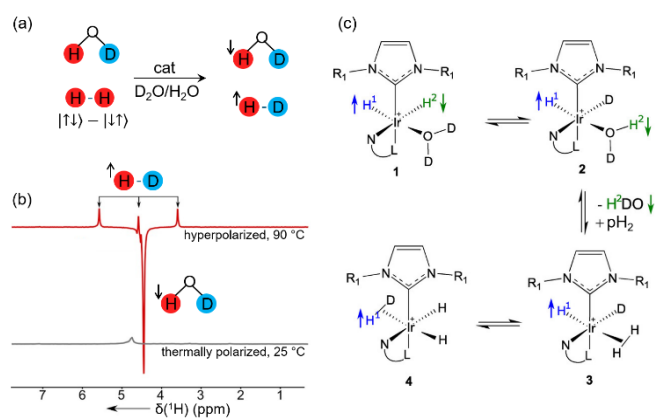

**Figure 7**. (a) H/D exchange, in the presence of a water-soluble iridium catalyst and histidine, leading to hyperpolarization of HD and HDO and (b) the resulting NMR spectrum, in comparison to the thermally polarized spectrum. (c) The proposed "NEPTUN" mechanism underpins the spectrum. *(a, b) Adapted from Lehmkuhl et al.* [148] *Copyright 2017 with permission from Wiley-VCH GmbH. License number 5851560561506. (c) Adapted from Emondts et al.* [89] *Copyright 2018 with permission from Wiley-VCH GmbH. License number 5851560116644.*

A NEPTUN-type mechanism was also suspected to be active in relayed hyperpolarization experiments where the goal was to further extend the SABRE hyperpolarization to heteronuclei in non-coordinating substrates like alcohols (e.g., methanol, ethanol) *via* proton exchange with a carrier amine.[149] The possible involvement of such a mechanism, in addition to OH/NH exchange, was inferred from magnetic field dependencies of $^{13}$C distortionless enhancement by polarization transfer (DEPT) signals, which, in addition to showing a peak at 6.5 mT, as expected for the conventional SABRE matching condition (relayed to the target *via* NH/OH exchange), there is an even more prominent peak after transfer at 19.2 mT, which is hypothesized to stem from the NEPTUN effect.[149] However, attempts to observe the hydride resonances indicative of NEPTUN directly were unsuccessful.

While most PHIP studies utilize dissolved organometallic catalysts, heterogeneous catalysis offers facile separation of the hyperpolarized products from the catalyst and can even be used in a packed-bed flow-reactor configuration.[150,151] Supported noble metals are amongst the most active hydrogenation catalysts. Unfortunately, only a tiny fraction of adducts (ca. 1-5%) are formed by pairwise addition, depending on nanoparticle size and reaction conditions. Rapid H adatom diffusion and facile exchange with gaseous H$_2$ conspire to destroy the singlet order in the nascent H adatom pair.[13,39,152,153] Intermetallic phases incorporating an active and inactive metal, such as Pt and Sn, allow for the tuning of molecular adsorption and diffusion dynamics through a combination of geometric and electronic effects.[154–156] Thus, as the fraction of Sn increases across the series Pt→Pt$_3$Sn→PtSn, the pairwise selectivity for hydrogenation of propene increases by more than three orders of magnitude.[154] After bubbling pH$_2$ through a D$_2$O suspension of Pt$_3$Sn@mSiO$_2$ (Pt$_3$Sn nanoparticles encapsulated in mesoporous silica) for 30 s, Zhao et al. observed hyperpolarization of the impurity protons of solvent molecules.[134] The effect was dubbed surface waters are magnetized from parahydrogen (SWAMP). Proton hyperpolarization in methanol-d$_4$ and ethanol-d$_6$ was also observed. The surface properties of Pt$_3$Sn now balance the necessary facile H$_2$ activation and suppression of diffusion.

The NEPTUN and SWAMP systems share a few similarities: (i) emission phase of the HDO peak. (ii) absence of signal enhancement at zero or high magnetic field, revealing a role of Zeeman interactions; (iii) monotonic growth of [HDO] with total $pH_2$ bubbling time; (iv) emergence of a dissolved HD (triplet) NMR signal. The last two are accounted for by the net isotope exchange reaction described in equation 3.

$$H_2 + D_2O \leftrightharpoons HD + HDO \tag{3}$$

For Pt surfaces, H/D exchange is mediated by reversible electron transfer from an H adatom to the metal and proton transfer to surface water to yield a hydronium-like species where H/D exchange occurs.[157] However, the oneH-PHIP/NEPTUN mechanism was only tentatively excluded by preliminary data revealing the different dependences of the SWAMP signals for exchangeable and non-exchangeable protons of $HOCD_3$ and $DOCHD_2$, respectively, on the total amount of H/D exchange. Plausible mechanisms are illustrated in **Figure 8a**.

While the monometallic Pt@$mSiO_2$ nanoparticles of Ref.[134] were found to be inactive as SWAMP catalysts, Norcott discovered that one could hyperpolarize water and methanol using a commercially available carbon-supported Pt nanoparticle catalyst when benzoquinone was added to the $D_2O$ suspension of the catalyst.[158] Maximum NMR signal enhancements (relative to thermal equilibrium at 1.4 T) approaching 45-fold were observed with ten equivalents of benzoquinone (w/w with respect to Pt/C). Benzoquinone is converted to hydroquinone during this process, which assists in increasing the turnover of fresh $pH_2$ on the surface and thereby increasing the level of polarization of methanol and water (**Figure 8b**).

While such signal enhancements are likely to increase with further catalyst development and optimization of experimental conditions, it remains to be seen whether the PHIP approach can rival the very high polarization levels achievable by dDNP for water.[140,141,143–146] As $pH_2$-based hyperpolarization techniques are inherently rapid, continuous, and low-cost, their use to achieve sufficient levels of water hyperpolarization could provide advantages to dDNP methods and significantly advance the application range of hyperpolarized water in biomedical research.

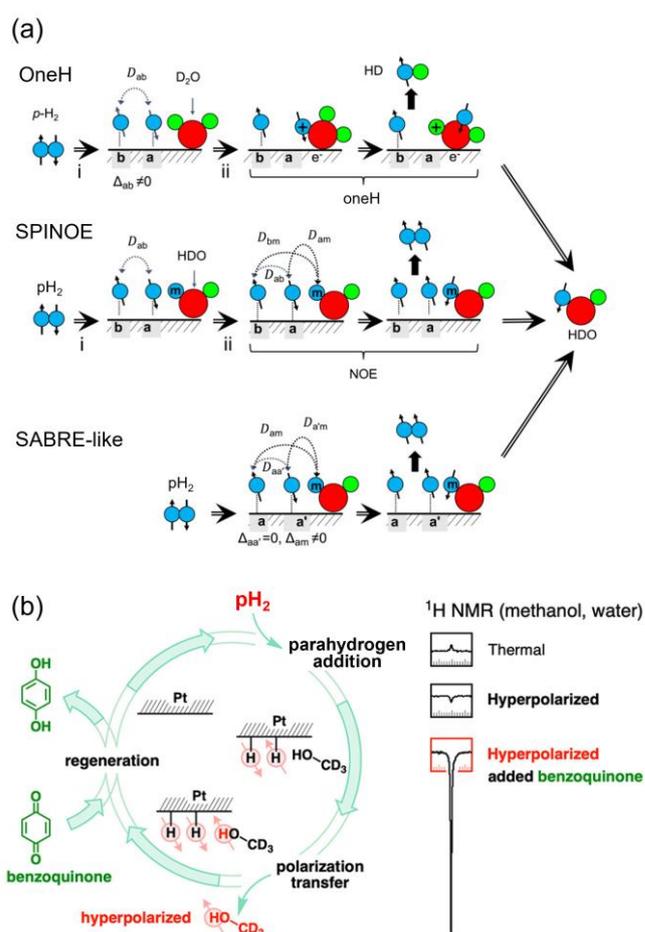

**Figure 8**. (a) Possible mechanisms underpinning surface-mediated hyperpolarization of liquid water. (b) Mechanism of benzoquinone scavenging of depolarized H adatoms on Pt/C, and its effect on the SWAMP signals. *(a) Adapted from Zhao et al.* [134] *Copyright 2018 with permission from Elsevier. License number 5851561162613. (b) Adapted with permission from Norcott.* [158] *Copyright 2023 The Author. Published by American Chemical Society. This publication is licensed under CC-BY-NC-ND 4.0.*

### 2.5. Metal-free PHIP: molecular tweezers and $pH_2$ activators

The chemical activation of $pH_2$ is crucial to derive enhanced NMR signals in PHIP. Commonly, transition metal catalysts are employed to mediate such activations and produce hyperpolarized substances. At the same time, the use of metal-free activators and catalysts for $pH_2$-based hyperpolarization, collectively named metal-free PHIP (MF-PHIP), is also documented.[111,129,130,159–166] This section focuses on several types of MF catalysts, their structures, hyperpolarization effects, and their unique mechanistic features (**Figure 9**).

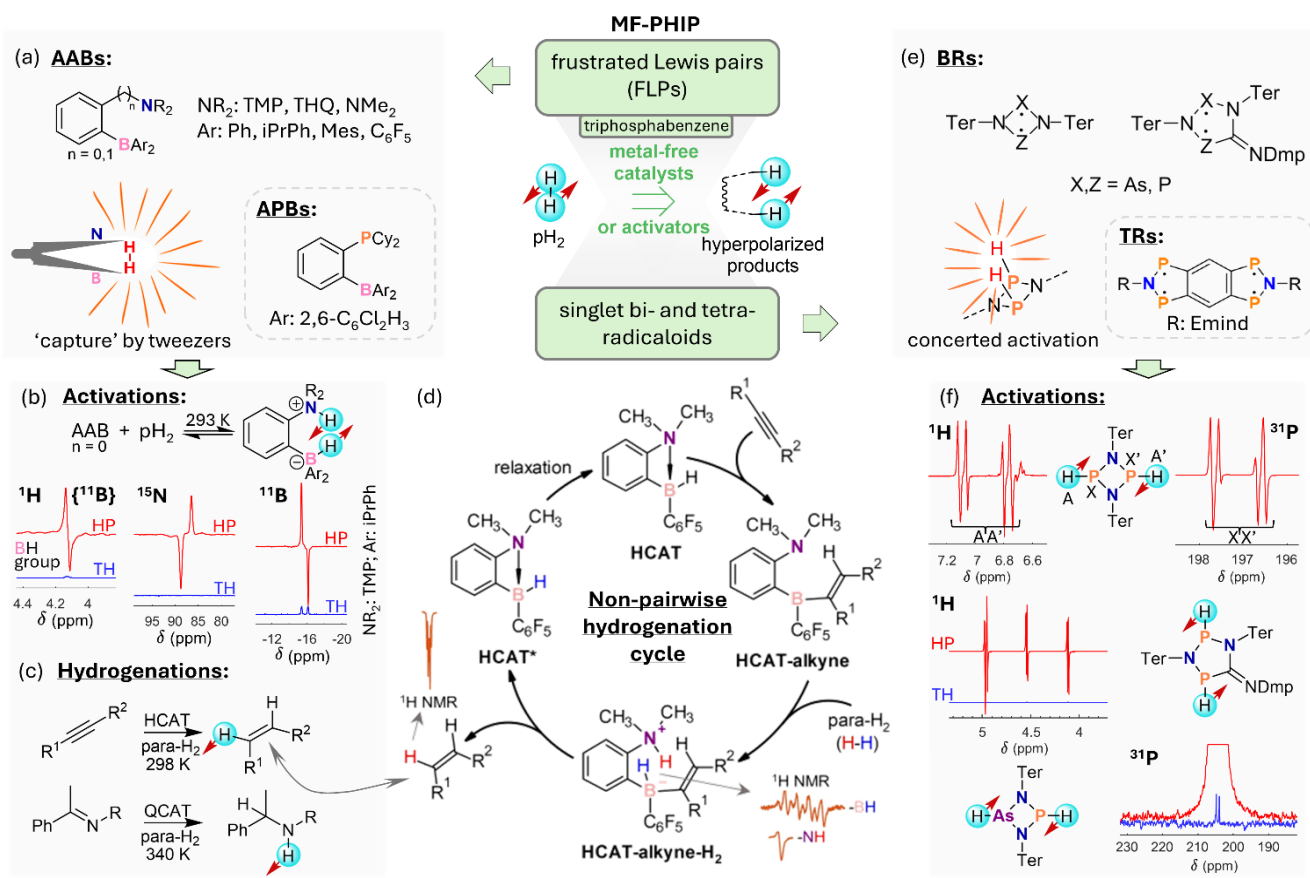

**Figure 9. An overview of metal-free PHIP (MF-PHIP).** The central top part highlights that MF-PHIP effects have been demonstrated in activation of pH$_2$ using frustrated Lewis pairs (FLPs) and using bi- and tetraradicaloids (BRs and TRs). Correspondingly, the structures of *ansa*-aminoborane (AAB) and *ansa*-phosphinoborane (APB) FLPs are presented in (a). AABs are referred to as molecular tweezers for pH$_2$. pH$_2$ activation under ambient conditions using ortho-phenylene AABs (n = 0) accompanied by the corresponding hyperpolarized NMR spectra is illustrated in (b). See (a) for the definition of n. In addition to $^1$H, $^{15}$N and $^{11}$B nuclei are also hyperpolarized spontaneously at high magnetic fields in this process. Alkyne and imine hydrogenation reactions catalyzed by AABs HCAT (n = 0; NR$_2$ = NMe$_2$; Ar = C$_6$F$_5$; R' = H) and QCAT (n = 1; NR$_2$ = THQ; Ar = C$_6$F$_5$), respectively, are shown in (c). Additionally, the catalytic cycle of alkyne hydrogenation using the HCAT AAB catalyst is presented in (d). Typical $^1$H NMR signals of the reaction intermediate and the reaction product are shown next to the corresponding structures in the cycle. Structures of BR and TR molecules that demonstrated hyperpolarization effects in pH$_2$ activations are depicted in (e). Corresponding examples of enhanced $^1$H and $^{31}$P NMR signals observed in reactions with BR molecules are shown in (f). The structures of the corresponding BR-H$_2$ adducts are depicted next to the spectra. Abbreviations: TMP = N-2,2,6,6-tetramethylpiperidinyl; THQ = N-tetrahydroquinolinyl; iPrPh = 2-isopropylphenyl; Mes = mesityl; Cy = cyclohexyl; Ter = 2,6-dimesitylphenyl; Dmp = 2,6-dimethylphenyl; Emind = 1,1,7,7-tetraethyl-3,3,5,5-tetramethyl-s-hydrindacenyl.

MF activations of $H_2$ are less common than those that rely on transition metal centers. They have recently attracted a lot of attention due to the possibility of using sustainable main-group elements to design less toxic and more environmentally friendly catalysts. MF catalysts for PHIP is an emerging research field that is still in its infancy. In this regard, frustrated Lewis pairs (FLPs)[167] are the most studied class of MF activators for $pH_2$. Specifically, various *ansa*-aminoborane (AAB) FLPs show pronounced hyperpolarization effects (**Figure 9a**). These compounds are referred to in the literature as 'molecular tweezers' that stretch but do not split $H_2$ molecules.[168] Recent studies revealed that the stretched H-H bond is relatively weak, making it possible to form various rotomeric forms in solution, including those with large H…H separations.[163] Nevertheless, AAB-$H_2$ adducts have motionally averaged *J*-coupling constants (2-4 Hz) between the $^1H$-$^1H$ pair, allowing for PHIP effects under high field (PASADENA) conditions.[111,163] Unlike homolytic oxidative addition to metal centers, AABs activate $pH_2$ heterolytically with a clear charge separation on the Lewis acidic boron and the Lewis basic nitrogen sites (**Figure 9b**), although the two protons remain spin correlated. Depending on the AAB structure, hyperpolarization of $^1H$, $^{11}B$, and $^{15}N$ can be observed in simple $pH_2$ bubbling experiments without harnessing dedicated pulse sequences.[36] Signal enhancements as large as 2000-fold at 9.4 T and room temperature have been observed in the resulting $^1H$ NMR spectra for the $pH_2$-originating protons of AAB-$H_2$. The size of this $^1H$ NMR signal gain depends strongly on the experimental conditions and is defined by relaxation and kinetic parameters.[163] Density functional theory (DFT) calculations have also revealed various conformational forms of AAB adducts and their transformations.

In addition to simple $pH_2$ activation, *ansa*-aminoboranes can be used in catalytic hydrogenations of alkynes[129] and imines[130] with $pH_2$ (**Figure 9c**). Interestingly, the catalytic cycles that lead to hyperpolarized akenes and amines are non-pairwise, implying that the $pH_2$-derived protons end up in different product molecules (**Figure 9d**). The hyperpolarization effects in this case are not expected to be observable in PHIP. However, due to the strong chemical shift anisotropy (CSA) of NH protons in the catalytic intermediates, a net negative polarization is generated from the $pH_2$ spin order through CSA-dipole-dipole cross-correlated relaxation.[130] For instance, in alkyne hydrogenations, this mechanism is revealed by the negative in-phase resonance of the NH group of HCAT-alkyne-$H_2$ intermediate. As this proton transfers to the final alkene product, a two-orders-of-magnitude enhanced negative signal of one of the added protons at the double bond of the resulting alkene appears in the $^1H$ NMR spectra at 9.4 T. This effect is related to oneH-PHIP in hydroformylation reactions catalyzed by metals (Section 2.3), but the underlying mechanisms of hyperpolarization, as well as chemical processes, are different. It is worth noting that the ability of cross-correlated relaxation to transform $pH_2$ spin order to a net polarization was also observed on metal complexes, e.g., by Aime et al. in $pH_2$ activations using Os and Ru clusters.[169,170]

AABs are reported to be generally water-intolerant, which is a significant obstacle to the wide application of MF-PHIP. However, *ansa*-phosphinoboranes (APBs) were demonstrated to show PHIP effects in the presence of several equivalents of $H_2O$ (**Figure 9a**).[162] Other FLPs showing hyperpolarization effects include Sn/P systems,[171] though it is strictly not a MF compound and will not be discussed here. In addition, aromatic triphosphabenzene was also shown to reveal hyperpolarization in the reaction with $pH_2$ at elevated temperatures (375 K).[159] However, the resulting $H_2$ adduct is prone to decomposition.

Singlet pnictogen radicaloids are another class of MF $pH_2$ activators that demonstrate prominent hyperpolarization effects. Electron spins in these open-shell molecules are coupled into a singlet state that does not possess free electron angular momentum, which excludes the deleterious influence of the radical centers on the nuclear spins. Typical examples include cyclic species with P and/or As radical centers isolated by surrounding bulky substituents for stabilization. This configuration maintains high reactivity towards small molecules, such as $H_2$, while preserving an open-shell structure. In the context of PHIP, biradicaloids (BRs) with four- or five-membered cycles are studied more extensively (**Figure 9e**), and any observed hyperpolarization effects strongly depend on BR symmetry.[165] With symmetric four-membered biradicaloids, $pH_2$ forms symmetric adducts. For instance, four-membered baricaloids form the AA'XX' spin system, which leads to enhanced $^1H$ and $^{31}P$ NMR signals in PASADENA experiments (**Figure 9f**).[164,165] Non-symmetric five-membered species form a system with weakly coupled protons, leading to only $^1H$ hyperpolarization. However, the transfer of $^1H$ hyperpolarization to $^{31}P$ can be achieved using ESOTHERIC NMR pulse sequences with $^{31}P$ NMR signal enhancements exceeding three orders of magnitude at 9.4 T.[165,172] Tetraradicaloids (TRs) are represented by a single example (**Figure 9e**),[166] which showed less pronounced but interesting hyperpolarization for the addition of the first and second equivalents of $pH_2$. Radicaloid systems are generally more reactive than FLPs, which makes them especially interesting for future developments that may lead to active catalysts, e.g., for hydrogenation or hydroformylation reactions.

MF-PHIP represents an exciting frontier in hyperpolarization method development. Using MF compounds such as FLPs and radicaloids introduces new mechanistic pathways and structural features that differentiate them from traditional metal-catalyzed systems. The unique hyperpolarization effects and mechanisms of MF-PHIP, including two-centered activation and non-pairwise hydrogen transfer, offer potential for innovative applications. Future research in this area promises to further enrich our understanding and utilization of these novel catalysts for hyperpolarization.

## 2.6. Revealing PHIP in subsequent chemical transformations

The involvement of hyperpolarized molecules in subsequent chemical transformations is an interesting application of PHIP that is nicely illustrated using metabolic reactions, such as pyruvate-to-lactate conversion[173] and fumarate-to-malate[174] conversions. Other examples include oxidation of hyperpolarized pyruvate with $H_2O_2$[175] or decarboxylation with yttrium polyaminoacarboxylate adducts,[176] methylation of N-heterocycles,[177] and conversion of hyperpolarized $^{15}NO_2^-$ via several reactions to make a range of products with enhanced $^{15}N$ NMR signals,[178] and other transformations.[179]

Typically, PHIP-enhanced NMR signals are visible without any subsequent reaction and can serve as a tool for kinetic measurements and identifying intermediates or additional products of the following reactions. However, if we consider symmetric molecules produced in reactions with $pH_2$, the product, although in a far from thermodynamic equilibrium nuclear spin state, may not exhibit observable hyperpolarization if the $pH_2$ addition site is at the center of symmetry. In this case, a subsequent symmetry-breaking chemical reaction(s) may be required to reveal otherwise unobservable hyperpolarization (**Figure 10a**). The quintessential example of this kind is the $pH_2$ molecule, which accommodates singlet nuclear spin order and does not yield NMR signals. As first shown by Bowers and Weitekamp,[15,16] one needs to break the symmetry of $pH_2$ in a chemical reaction to observe hyperpolarization. Notably, symmetric molecules can typically host long-lived spin orders.[180,181] Such molecules can be synthesized using $pH_2$ in hydrogenative and non-hydrogenative reactions. For example, the syntheses of ethylene from acetylene and $pH_2$ (**Figure 10b**), dimethyl maleate (DMM) from dimethyl acetylene dicarboxylate (DMAD) (**Figure 10c**), and para-$^{15}N_2$ using SABRE were described.[182–189]

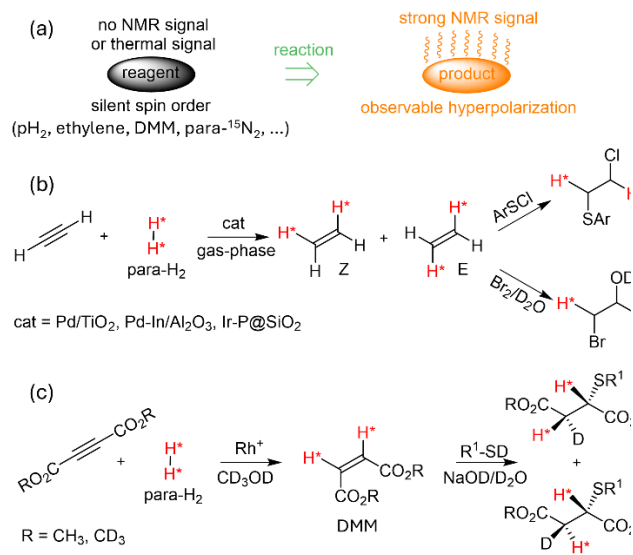

**Figure 10**. (a) A general scheme of formation of compounds with hyperpolarized spins in chemical reactions of symmetric molecules, such as $pH_2$, ethylene, DMM, and para-$^{15}N_2$, accommodating otherwise unobservable nuclear spin orders. (b) The chemical synthesis of Z- and E-ethylene from acetylene and $pH_2$ over various heterogeneous catalysts, as well as subsequent reactions revealing hyperpolarization. (c) The chemical synthesis of dimethyl maleate (DMM) molecules in hydrogenation with $pH_2$ over a cationic $Rh^+$ catalyst, followed by their subsequent reaction with thiol molecules to reveal hyperpolarization. Abbreviations: DMM - dimethyl maleate.

The case of ethylene is fundamentally interesting since it, like $H_2$, has nuclear spin isomers of molecules (NSIMs) that differ by rotational and spin degrees of freedom due to the coupling of nuclear spin and rotational states through the symmetry properties of their respective wave functions.[190] Briefly, there are four NSIMs for ethylene that can be classified according to the symmetries of the nuclear spin state for the $D_{2h}$ molecular point group using Mulliken symbols: $A_g$ (one quintet and two singlets), $B_{1u}$ (triplet), $B_{2u}$ (triplet), and $B_{3g}$ (triplet). Depending on the stereoselectivity of the hydrogenation of acetylene, syn and anti $pH_2$ addition products, Z- and E-ethylene, respectively, can be produced (**Figure 10c**),[182] which is primarily determined by the hydrogenation catalyst employed. For instance, supported Pd nanoparticles are less selective and produce both Z- and E-ethylene products,[182] whereas immobilized complexes of Ir are more selective, leading primarily to Z-ethylene.[185] Interestingly, the subsequent reactions of ethylene produced using different catalysts with sulfenyl chlorides[182] or $Br_2/D_2O$[185,191] reveal different lifetimes of the non-equilibrium spin states in ethylene. This was rationalized based on the interconversion between different NSIMs of ethylene,[184] providing insights that in the gas phase, Z-ethylene has only one long-lived component. In contrast, E-ethylene has two long-lived components due to the imbalance of NSIMs with different inversion symmetry in the latter case.[182,185] Lifetime constants of more than 15 min were measured by unlocking the hyperpolarization in the subsequent reactions for gaseous E-ethylene.

Another example of revealing the latent polarization inherited from $pH_2$ is documented for the reaction of thiols with DMM produced from DMAD in a liquid-state hydrogenation over a Rh(I) cationic catalyst (**Figure 10c**).[183] As in the case of ethylene, storage of the non-equilibrium nuclear spin order after the hydrogenation was demonstrated in this study. The thiol reaction allowed for the lifetime measurement of the populated long-lived singlet spin order at a high field of up to 4.7 min. Interestingly, in this case, the unique symmetry of the molecule, providing slight magnetic inequivalence of the added proton pair in DMM, also allowed alternative ways to reveal hidden spin state populations that do not require chemical transformation. Instead, such hidden spin states can be converted into observable magnetization using magnetic field cycling[192] or applying RF fields.[193–195] These methods do not apply to ethylene, as all its protons are magnetically equivalent.

In addition to ethylene and DMM, para-$^{15}N_2$ has also been reported to form from SABRE-hyperpolarized $^{15}N$-labeled tetrazine[188] and diazirines[189] in chemical reactions involving these agents. However, the successful formation of para-$^{15}N_2$ in these cases was inferred only from the absence of a $^{15}N_2$ signal in $^{15}$N NMR spectra after its production step. Similarly to pH$_2$, para-$^{15}$N$_2$ is NMR silent. So far, no subsequent symmetry-breaking reaction of para-$^{15}$N$_2$ that reveals its singlet spin order has been reported to our knowledge.

Overall, the availability of methods that use pH$_2$ to produce symmetric molecules with long-lived nuclear spin orders can expand the range of chemical reactions that can be studied by providing flexible time windows. For example, it can allow one to go beyond classical hydrogenation into chemically inequivalent sites or H$_2$ activation, in the case of pH$_2$, in electrophilic additions to double bonds, as in the case of ethylene and DMM. Furthermore, the analysis of the generated spin order lifetimes in molecules such as ethylene and para-$^{15}$N$_2$ can provide essential insights into the fundamentals of NSIMs and underlying molecular physics.

## 2.7. Spreading hyperpolarization *via* chemical exchange: PHIP-X and SABRE-RELAY

In recent years, a new approach has emerged to boost the NMR signal for molecules that do not contain functionalities suitable for direct PHIP or SABRE (natively or on a side arm). Two versions of this approach are based on reversible proton exchange between a hyperpolarized carrier and a to-be-hyperpolarized molecule and are termed PHIP by chemical exchange (PHIP-X)[196] and SABRE-Relay.[197] In both methods, traditional PHIP or SABRE is used to polarize a 'transfer' or 'carrier' molecule, whose polarization is then transferred to a secondary target molecule *via* the exchange of hyperpolarized OH/NH protons (**Figure 11, 12a**). These approaches have allowed a significant expansion of the substrate scope of PHIP and SABRE in recent years.

In the case of PHIP-X[196] (also referred to as PHIP-Relay),[198] propargyl alcohol, propiolic acid, or propargyl amine were hydrogenated with pH$_2$ using a homogeneous Rh catalyst in an aprotic solvent, such as acetone, to produce a hyperpolarized 'transfer' agent (**Figure 11**).[196] Strong spin-spin interactions distribute the polarization among the protons of the transfer agent, including the labile OH (or NH) proton. The polarization of this labile proton is relayed to the spin system of a third molecule by chemical exchange, where, again, spin-spin couplings, low magnetic field, or RF spin order transfer sequences (RF-SOT) facilitate the transfer of the polarization to other nuclei such as $^1$H, $^{13}$C, or $^{15}$N.[198,199] PHIP-X was shown to spontaneously polarize labile protons to ca. 0.4% for ethanol and water, 0.07% for lactic acid, 0.005% for pyruvic acid, and at least 0.009% $^{13}$C polarization for glucose. Using RF-SOTs, 1.2% $^{15}$N polarization was achieved for urea, where the $^{15}$N coupling to the labile proton is large, 0.024% for $^{13}$C glucose[198], 0.026% for $^{13}$C lactate[199] and ca. 0.007% for $^{13}$C methanol.[199] The balance here is reached when proton exchange is slow enough to allow the *J*-coupling to transfer polarization to the labile proton of the carrier first and then from the labile proton to other nuclei of the target. At the same time, the exchange must be fast enough such that spin relaxation would not destroy polarization before the target is polarized. Therefore, even higher polarization values are expected to be achievable after thoroughly tuning the exchange parameters. Also, the polarization transfer is faster and more efficient when transferred to nuclei directly bound to the labile proton, such as $^{15}$N or $^{13}$C, with the strongest *J*-coupling constants. An advantage of using PHIP to hyperpolarize the transfer agent compared to SABRE is that the molecule can be polarized up to unity by the direct addition of pH$_2$; a disadvantage is that it is irreversible as the addition step can be performed only once (one addition of pH$_2$ per transfer agent).

In SABRE-Relay, classical SABRE is used to hyperpolarize ligating carriers, typically NH$_3$ and amines, with secondary NH/OH exchange effectively relaying polarization to non-ligating target substrates.[200,201] Consequently, SABRE-Relay has been used to hyperpolarize alcohols,[149,202] sugars,[203] silanols,[61] lactate esters,[204] natural products,[205] and many other functional groups[197,206] that do not interact with the SABRE catalyst directly. As SABRE-Relay, like PHIP-X, depends on transferring proton magnetization, direct polarization of heteronuclei (analogous to SABRE-SHEATH) is impossible. However, the polarization of the exchanging OH group allows the polarization of heteronuclei either spontaneously (i.e., by free evolution), or with the help of RF-SOT. So far, SABRE-Relay has achieved 2.6% $^1$H,[202] 2.3% $^{29}$Si,[61] 1.1% $^{13}$C,[202,203] 0.2% $^{19}$F,[202] and 0.04% $^{31}$P[202] polarization levels. These NMR signal enhancements are generally lower than typical values achieved by conventional SABRE because the relayed polarization is derived from a finite carrier polarization affected by spin relaxation during the chemical exchange. Current studies have optimized factors such as amine type and concentration ratios to increase the target polarization.[202,204,205]

A major limitation of PHIP-X and SABRE-Relay is that it cannot be performed in alcohol or aqueous solvents as their exchangeable protons will compete with the ones of the target molecule. Accordingly, it is commonly performed in dry dichloromethane or chloroform, which may pose a significant challenge for insoluble substrates in these media. SABRE-Relay has already shown potential in molecular analysis. It can enhance sugars[203] and natural products[205] at concentrations as low as tens of micromolar with a single NMR scan. Notably, it can give single-scan quantification of isomeric ratios for OH-containing molecules, such as α and β forms of glucose and fructose (**Figure 12b**)[203] or diastereomers of natural oils like (−)-carveol (**Figure 12c**).[205] Thanks to the continuous nature of polarization production in SABRE, relayed polarization transfer is expected to become better understood, improved to increase polarization levels, and applied to an ever-increasing scope of target molecules in the years ahead.

Compared to direct pH$_2$ addition (PHIP-X), using reversible exchange (SABRE) to polarize the transfer agent has the advantage that the transfer agent can be continuously re-polarized; a disadvantage is that the polarization yield is usually lower. Despite recent progress,[199,204,207] the detailed and quantitative description of chemical exchange and the spread of polarization within the target is still not fully understood.

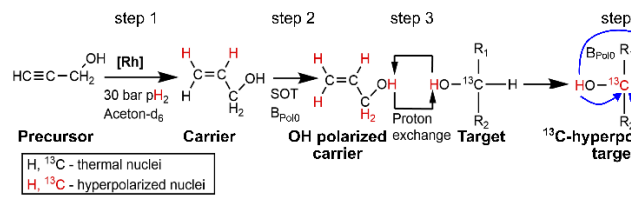

**Figure 11. Schematic view of parahydrogen-induced polarization by chemical exchange (PHIP-X).** PHIP-X consists of four essential steps: hydrogenation of the carrier agent (step 1), the polarization of the exchanging protons (step 2), transfer of the exchanging protons from the carrier to the target molecule (step 3), and polarization of the target nucleus (step 4) using RF-induced spin order transfer technique or free evolution at low and ultralow magnetic fields. *Adapted with permission from Them et al.*[199] *Copyright 2024, The Author(s). This publication is licensed under CC BY 4.0.*

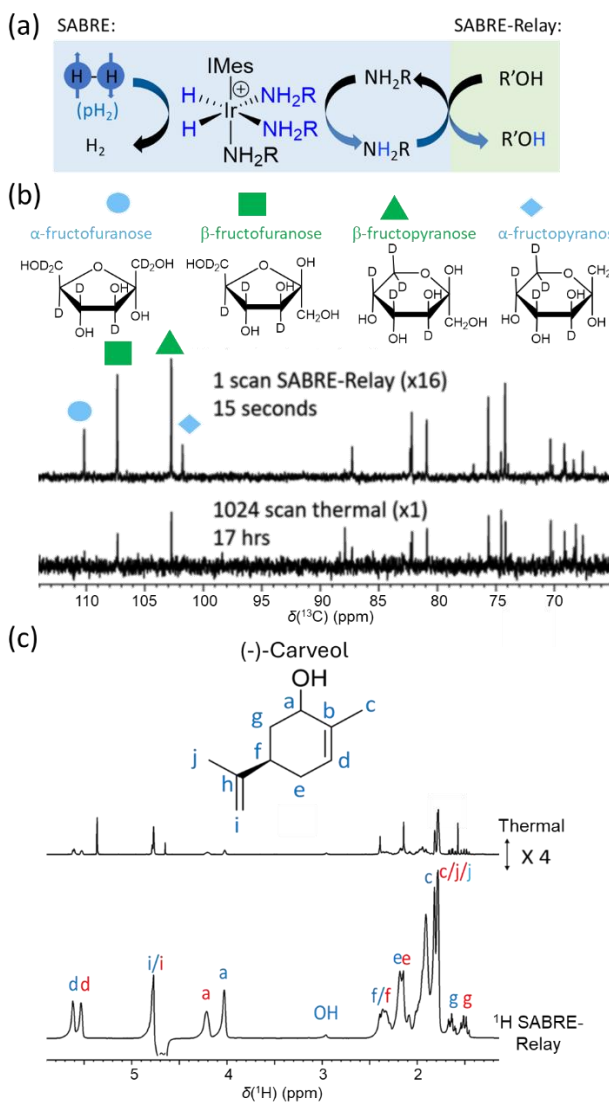

**Figure 12. Demonstration of SABRE-Relay effect.** (a) Depiction of the SABRE-Relay method. (b) SABRE-Relay can allow quantification of isomer ratios of fructose in a single scan $^{13}C$ measurement or (c) quantification of diastereomer ratios of the natural product (−)-carveol from a single scan $^1H$ measurement. *Details for (b):* $^{13}C\{^1H\}$ NMR spectra (right) acquired for 40 mM of D-fructose (natural $^{13}C$ abundance) with 23.8 mM benzyl-$d_7$-amine and 4.8 mM of [Ir(Cl)(COD)(SIMes-$d_{22}$)] (where COD is cis,cis-cyclooctadiene and SIMes is 1,3-Bis(2,4,6-trimethylphenyl)-4,5-dihydroimidazol-2-ylidene) in a 0.65 mL DCM-$d_2$ : DMF (1.6 : 1) mixture measured at 9.4 T. The bottom spectrum shows the result of a thermally polarized signal averaging over 1024 scans (approx. 17 h), and the middle spectrum represents the single scan SABRE-Relay hyperpolarization measurement recorded after shaking the sample with p$H_2$ at 6.5 mT. *Details for (c):* Exemplar single scan thermally polarized (above) and $^1H$ SABRE-Relay hyperpolarized (lower) $^1H$ NMR spectra for a sample of [IrCl(COD)(IMes)] (5 mM), $NH_3$ (30 mM), (−)-carveol (25 mM) and p$H_2$ (3 bar) in DCM-$d_2$ (0.6 mL). The resonance labels in red and blue correspond to the two diastereomers. The Hyperpolarized NMR spectrum is recorded immediately after shaking the sample for 10 s with fresh p$H_2$ at 6.5 mT. *(a, c) Adapted with permission from Alshehri et al.*[205] *Copyright 2023 Royal Society of Chemistry. This publication is licensed under CC BY-NC 3.0. (b) Adapted with permission from Richardson et al.*[203] *Copyright 2019 Royal Society of Chemistry. This publication is licensed under CC BY-NC 3.0.*

## 2.8. Perspectives: PHIP in enzymatically catalyzed reactions

The use of p$H_2$ and hydrogen-deuterium scrambling has been explored in a non-NMR context for several decades to study kinetics and intermediates of enzymes thanks to the slow conversion of p$H_2$ to o$H_2$ in pure water of about tens to hundreds of minutes.[125] Such scrambling can thereby help to understand the chemisorption and exchange process by forming HD based on a hydrogen and deuterium source.[208] Detection of o$H_2$ formation after supplying p$H_2$ yields information on the splitting and recombination of $H_2$.[209] This information can then give a clearer kinetic picture of hydrogen-activating enzymes.[210] The particular focus, therefore, is on hydrogenases,[208–211] an important class of enzymes involved in the hydrogen activation process of, e.g., anaerobic organisms.[212–215] Hydrogenases promise to be blueprints for eco-friendly catalysts for hydrogen activation on the way to produce green energy.[216] Therefore, understanding the detailed hydrogen activation of hydrogenases could facilitate the design of potent eco-friendly catalysts. Three different types of hydrogenases are currently known: [FeFe]-hydrogenases, [NiFe]-hydrogenases, and the [Fe]-hydrogenase.[217] While reaction intermediates of the first two hydrogenases could be well characterized by available methodologies such as EPR, X-ray diffraction, and IR, investigating the [Fe]-hydrogenase has proven to be more challenging. This is because the iron center of the [Fe]-hydrogenase is $Fe^{II+}$ encapsulated in a guanylyl pyridinol (FeGP) and remains in a diamagnetic state through the whole catalytic cycle (**Figure 13a**). Under catalytic conditions, methenyl-tetrahydromethanopterin (methenyl-$H_4MPT^+$) is bound by the protein, bringing together FeGP and methenyl-H$_4$MPT$^+$. This is the active site that heterolytically cleaves molecular hydrogen into a proton and a hydride, whereby the hydride is stereo-specifically transferred to the H*pro*-R position of the methylene carbon of methylene-H$_4$MPT[218] (**Figure 13b**). So far, computational models have predicted several iron-hydrogen species in the catalytic cycle, none of which could be experimentally verified. The use of pH$_2$ in hyperpolarization experiments has recently changed this.[132] When pH$_2$ was supplied to an aqueous buffer containing [Fe]-hydrogenase and methenyl-H$_4$MPT$^+$, the appearance of a hyperpolarized PNL (Section 2.2) as well as HD and HDO NEPTUN PHIP signals (Section 2.4) was observed (**Figure 14a**).[132] The former state is created when pH$_2$ reversibly binds to the enzyme. In addition, hyperpolarized HD signals, first observed with iridium catalysts in the context of the NEPTUN effect,[219] were also observed in the presence of hydrogenase when the buffer was partially deuterated. This finding further indicates an isotope exchange with the solvent. In addition, rapid exchange between an enzyme-bound ensemble of hydrogen where both hydrogens are distinguishable, and a state where both hydrogens are indistinguishable on the NMR time scale, needs to occur. An estimate for the lifetime range of 1-100 µs was found, and chemical shifts and $^1$H-$^1$H *J*-coupling constants between these hydrogens were estimated. Optimized structural models based on the X-ray crystal structure of the hydrogenase allowed for the computation of $^1$H chemical shifts and $^1$H-$^1$H *J*-coupling constants for the hydrogen atoms within the active site and correlation with the experimental observations. Considering all this, an intermediate could be identified that had only been predicted previously (**Figure 14b**).[220] The optimized structure reveals the presence of an iron hydride and the involvement of the oxopyridine site during the activation process. The determined intermediate supports the previously theoretically predicted process during which an oxo-pyridine moiety serves as a base for hydrogen activation during the catalytic process.

With this demonstration, the use of pH$_2$-enhanced magnetic resonance has evolved into a tool to study the biochemistry and catalysis of hydrogenases and to investigate so far undetectable reaction intermediates of this important class of enzymes. It is envisioned that the concept can be used to study additional hydrogen-activating enzymes, such as the other two types of hydrogenases and potentially nitrogenases.

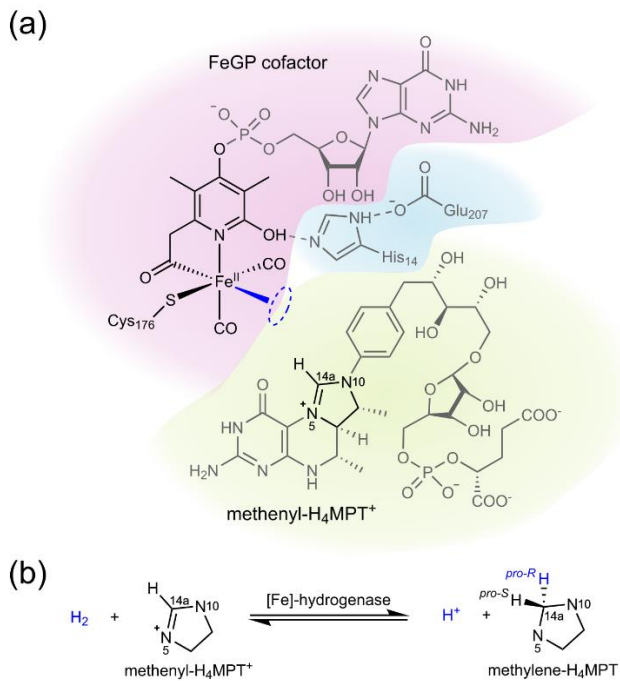

**Figure 13. The active site of the mono iron hydrogenase and the hydrogenation reaction** (a) The active site of the [Fe]-hydrogenase, including the iron guanylyl pyridonol and methenyl-H$_4$MPT+. (b) Hydride transfer to the H*pro*-R position of the substrate forming methylene-H$_4$MPT. *The figure is adapted from Ref.* [132].

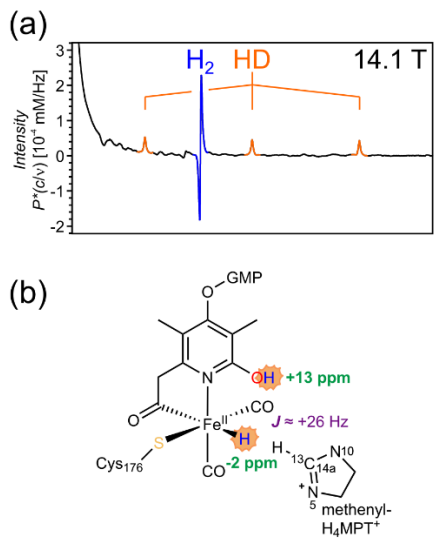

**Figure 14. Using PHIP to reveal hitherto unobservable intermediates of [Fe] hydrogenase.** (a) Hyperpolarized signals were observed when pH$_2$ was supplied to an enzymatic solution of the hydrogenase containing methenyl-H4MPT+. A PNL effect and oneH-PHIP hyperpolarized HD signals demonstrate the activation of pH$_2$ by hydrogenase. (b) The reaction intermediate was determined using observed NMR parameters and analysis of chemical exchange, and it was derived from a crystal structure quantum mechanical optimization of the structure. *The figure is adapted from Ref.* [132].

## 3. Outlook

Almost forty years after its first discovery, PHIP remains a source for ever-surprising novel applications and unique revelations, some of which are discussed in this review. In the wake of biomedical hyperpolarized MRI, recent discoveries of efficient ways to hyperpolarize $^{13}C$-labeled pyruvate led to in vivo imaging.[68,69,71,72] The high levels of polarization achieved using PHIP led to observations of effects such as radio amplification by stimulated emission of radiation (RASER) and spin diffusion in the solid state that attracted attention beyond the chemical community.[47,221–227] Continued exploration and development of PHIP has and will likely continue to yield tools and techniques that complement more traditional methods, which have their shortcomings.

One promising direction lies in the continued development of catalysts and reaction conditions that enable PHIP with catalytic systems previously considered incompatible with PHIP. The ability to perform PHIP in these settings can lead to more sustainable and versatile applications. In a way toward this, the field of applied catalysis will be enriched among others through synergetic development with PHIP as it was when the homogeneous catalysts were tuned to reach exclusive *trans* hydrogenation for a direct hyperpolarization of fumarate,[33] or when PHIP in homogeneous,[15–17] heterogeneous,[13,31,32,228,229] and metal-free[129,130,160] systems was demonstrated. Heterogeneous PHIP is currently progressing toward in vivo lung imaging.[36,37]

It is worth highlighting the innovative combination of hyperpolarization for boosting NMR sensitivity while maintaining its quantitative nature. Tessari and others proposed two compelling approaches to utilize SABRE in analytical chemistry. In one method, analytes are introduced and hyperpolarized via SABRE, enabling concentration evaluations with near-nanomolar sensitivity.[230–234] This demonstrates the potential of SABRE as a powerful quantitative tool. The second approach focuses on hyperpolarizing the hydride signals of the catalyst, exploiting the chemical shifts induced by the coordination of trace analytes to the Ir-complex.[84,10,235–238] This method transforms the hyperpolarized hydrides into dynamic probes for chemical analysis, opening avenues for studying even minute concentrations of analytes with precision.

Furthermore, integrating PHIP with advanced NMR techniques, such as ultrafast sequences and ultra-low and high-field instrumentation, may allow for detecting and studying transient intermediates and reaction dynamics with unprecedented sensitivity, and spectral and temporal resolution.[239–243] This can significantly enhance our understanding of complex chemical processes, lead to the discovery of new reaction pathways, and uncover details of catalytic mechanisms.

All discussed here was made possible by utilizing the correlation of spins in $pH_2$ and its effects on the nuclear spin polarization of other interacting neighboring atomic nuclei. Observing non-equilibrium polarization with NMR provides chemical resolution, enabling chemical analysis on an atomic level. NMR, however, is relatively slow, although, in some cases, like in photo-PHIP, the time resolution was boosted while preserving the chemical resolution. For example, a similar boost in time resolution was achieved when photo-induced radical pairs could be generated with a short UV irradiation, resulting in time-resolved chemically induced dynamic nuclear polarization.[244] Many methods discussed here are still under development and require significant expertise that currently limits their wider use. However, it is worth being aware of such effects, as, for example, they could provide invaluable information regarding the mobility of $H_2$ on catalytic centers that are inaccessible to other methods, as was exemplified here with hydrogenase studies.

## Acknowledgements

ANP, JBH acknowledge support from German Federal Ministry of Education and Research (BMBF) within the framework of the e:Med research and funding concept (01ZX1915C, 03WIR6208A hyperquant), DFG (555951950, 527469039, 469366436, HO-4602/2-2, HO-4602/3, HO-4602/4, EXC2167, FOR5042, TRR287). MOIN CC was founded by a grant from the European Regional Development Fund (ERDF) and the Zukunftsprogramm Wirtschaft of Schleswig-Holstein (Project no. 122-09-053). VVZ is grateful for the support from the Research Council of Finland (grant number 362959) and the University of Oulu (Kvantum Institute). SBD and BJT acknowledge the UK Research and Innovation (UKRI) under the UK government's Horizon Europe funding guarantee (grant number EP/X023672/1). GB gratefully acknowledges financial support from the DFG (BU 911/22-2). CRB acknowledges NSF grants CHE-2108306, CBET-1933723, and the National High Magnetic Field Laboratory User Collaborative Grants Program, which is supported by NSF DMR- 2128556 and the State of Florida. SG acknowledges funding from the DFG (SFB1633 project B03) and the Max Planck Society.

**Keywords:** parahydrogen • catalysis • hyperpolarization • mechanisms • NMR